\documentclass[runningheads]{llncs}
\usepackage{graphicx}
\usepackage{booktabs}
\usepackage[table]{xcolor}
\usepackage[colorlinks]{hyperref}
\usepackage{sidecap}
\usepackage[misc]{ifsym}
\usepackage{enumitem}
\usepackage{subfig}
\usepackage{pgfplots}
\usepackage{wrapfig}
\pgfplotsset{width=10cm,compat=1.5}
\usepackage[symbol]{footmisc}

\begin{document}
\title{COPD Classification in CT Images Using a 3D Convolutional Neural Network}
\titlerunning{COPD Classification Using a 3D CNN}
%
\author{Jalil Ahmed\inst{1}(\Letter)  \and
Sulaiman Vesal\inst{1} \and Felix Durlak \inst{2} \and Rainer Kaergel \inst{2} \and\\
Nishant Ravikumar\inst{1,3} \and{Martine R\'emy-Jardin}\inst{4}  \and Andreas Maier\inst{1}}

\authorrunning{Ahmed et al.}
%
\institute{Pattern Recognition Lab, Friedrich-Alexander-Universit\"at Erlangen-N\"urnberg, Germany\\ 
\and Siemens Healthcare GmbH, Forchheim, Germany\\
\and CISTIB, Centre for Computational Imaging and Simulation Technologies in Biomedicine, School of Computing, University of Leeds,\\ 
United Kingdom\\
\and  CHRU Lille, D\'epartement d'Imagerie Thoracique, Lille, France\\
\email{jalil.ahmed@fau.de}}
\maketitle              
\begin{abstract}
Chronic obstructive pulmonary disease (COPD) is a lung disease that is not fully reversible and one of the leading causes of morbidity and mortality in the world. Early detection and diagnosis of COPD can increase the survival rate and reduce the risk of COPD progression in patients. Currently, the primary examination tool to diagnose COPD is spirometry. However, computed tomography (CT) is used for detecting symptoms and sub-type classification of COPD. Using different imaging modalities is a difficult and tedious task even for physicians and is subjective to inter-and intra-observer variations. Hence, developing methods that can automatically classify COPD versus healthy patients is of great interest. In this paper, we propose a 3D deep learning approach to classify COPD and emphysema using volume-wise annotations only. We also demonstrate the impact of transfer learning on the classification of emphysema using knowledge transfer from a pre-trained COPD classification model.

\end{abstract}
\section{Introduction}
COPD is a lung disease characterized by chronic obstruction of lung airflow that interferes with normal breathing, and it is not fully reversible \cite{WHO}. It is considered to be the 4\textsuperscript{th} leading cause of death by 2030 \cite{10.1371/journal.pmed.0030442}. Generally, COPD is caused by a mixture of two sub-types, emphysema and small airway disease (SAD). Emphysema is the permanent abnormal enlargement of air spaces along with the destruction of their wall without prominent fibrosis. Emphysema is further classified into panlobular, centrilobular, and paraseptal emphysema based on the disease distribution in the lung, shape, and location of the affected area using lung CT images. The pulmonary function test is the gold standard for COPD detection. However, limitations in early detection and the ability of radiographic studies to visualize COPD sub-types have made CT a recent focus in the diagnosis and categorization of COPD. COPD is often misdiagnosed in early stages because of the similarity of initial symptoms to common illnesses, and lack of significant symptoms until the advanced stage \cite{doi:10.1111/j.1742-1241.2007.01447.x}. The shortcomings in the diagnosis are addressed using computer-aided detection (CAD). CAD techniques allow for a reduction in misdiagnosis rate leading to possible prevention of disease progression, complications, improving management, and early mortality.  \par
Automated COPD classification is conventionally approached using traditional machine learning techniques. Recently, deep neural networks are also employed for COPD classification in CT images \cite{Hatt2018}. The majority of studies use some prior information such as landmark slices \cite{Gonzalez2018}, regions of interest (ROIs) \cite{Ersa2015}, and meta-data \cite{Ying2016} in cooperation with 2D neural networks. In a 2D convolutional neural network (CNN), slices from the CT volume are used as input samples. This leads to inherent failure to leverage context from adjacent slices. These methods also require slice-wise annotations, which is a time-consuming and tedious task. We hypothesize that a solution to this problem is to use a 3D CNN which could overcome limitations for COPD classification using 3D convolutional kernels and volume-wise annotations. A 3D CNN can extract larger spatial context to preserve more discriminative information which subsequently could improve COPD classification.\par
In this study, we propose a 3D CNN for the volumetric classification of COPD as well as emphysema. We first train our proposed model for the classification of COPD vs. healthy. We then investigate the effects of transfer learning by using features learned during COPD classification and fine-tune the model for emphysema classification.

\section{Methods}
\begin{figure}[!t]
	\begin{minipage}[]{\columnwidth}
		\makeatletter
		\makeatother
		\centering
		\subfloat[]{%
			\includegraphics[width=0.19\textwidth, angle=90]{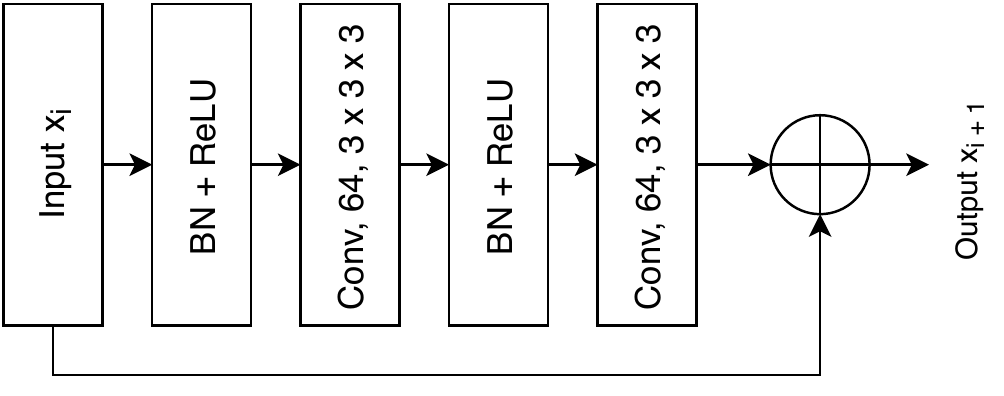}%
			\label{fig:voxres_module}
		}\\
		\subfloat[]{%
			\includegraphics[width=\textwidth]{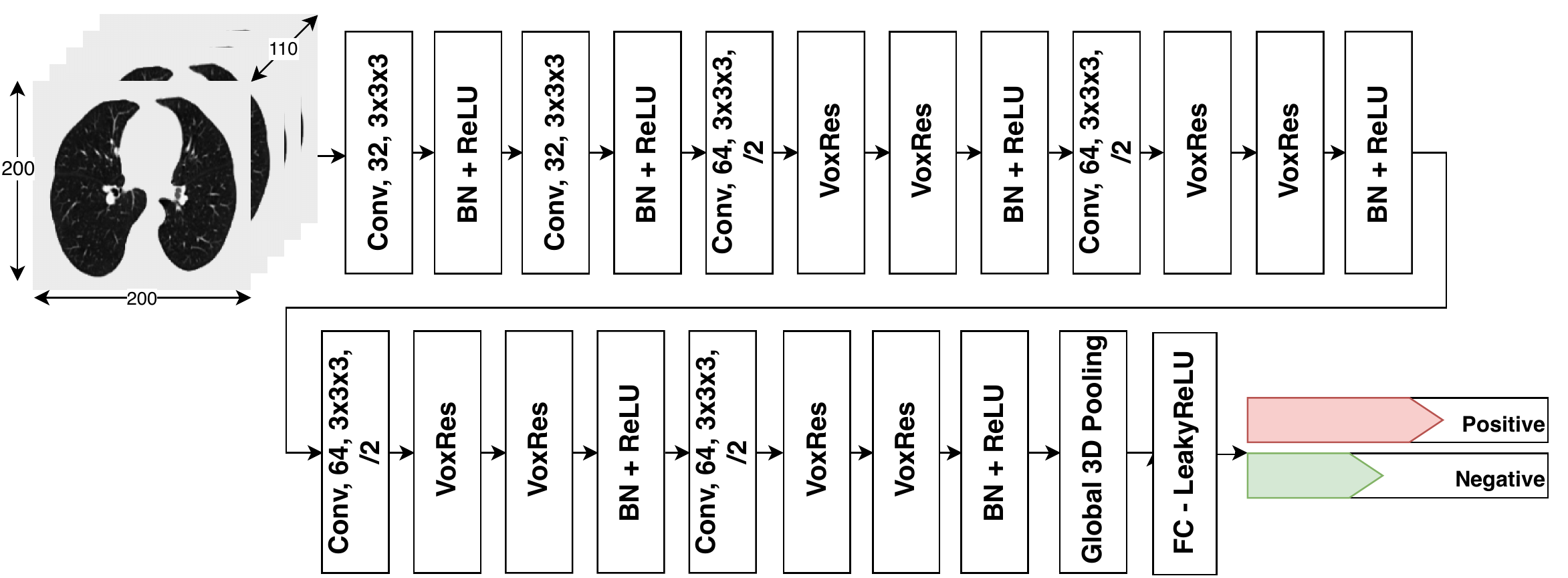}%
			\label{fig:voxresnet_segmentation}%
		} 
		\caption[VoxResNet]{(a) An illustration of VoxRes module (b) the proposed VoxResNet variant.}
		\label{fig:voxresnet_paper}
	\end{minipage}
\end{figure}

We utilized VoxResNet \cite{Chen_et_all} which extends deep residual learning to  3D volumetric data. The VoxRes module introduces identity mapping by adding input features with a residual function using a skip connection (Fig. \ref{fig:voxres_module}). We modified the VoxResNet by removing the auxiliary classifiers and adding fully connected layer for classification task. All the operations in VoxResNet are implemented in 3D to strengthen the volumetric feature representation learning. Fig. \ref{fig:voxresnet_paper} shows the proposed network architecture. The 2 initial convolutional layers have 32 filters each. Further layers consist of 64 filters each. Instead of a conventional pooling layer approach, after every 2 VoxRes modules, a convolutional layer with stride 2 was added to reduce the feature map size. Each convolutional layer uses a kernel size of \(3 \times 3 \times 3\) and is followed by batch normalization and a ReLU activation function. A 3D global-pooling layer with 64 units is used to preserve the 3D information of the last feature map. Afterward, a fully connected layer with the leaky-ReLU activation function is added. We also used an L2-regularizer to prevent sharp learning spikes and achieve a smooth learning curve. The network is implemented in the Keras/TensorFlow framework. 

\subsection{Data}
\label{sec:data}
We trained the proposed model using CT images from two data sets called SI-I and SI-II. SI-I consisted of inspiration scans with the soft kernel. All the subjects were annotated into different categories based on the Global Initiative for Chronic Obstructive Lung Disease (GOLD) system. We marked all the instances with GOLD stage 0 as healthy and those with GOLD stage 1-4 as COPD cases. \par
SI-II data set consists of 1000 patients with at least 2 scans per patient. We used the inspiration scan with the soft kernel. SI-II was annotated for emphysema and its potential multiple sub-types after visual assessment by medical experts. The non-emphysema label included patients who were healthy or had respiratory diseases other than emphysema. Sub-types of emphysema are not mutually exclusive. Both data sets were imbalanced. The SI-I data set has a ratio of 3/4 for the present annotations healthy vs. COPD.
SI-II consisted of 572 non-emphysema samples and 366 emphysema samples. The emphysema sub-types distribution was also imbalanced (panlobular: 4, centrilobular: 96, paraseptal: 32, multiple sub-types: 233).

\begin{figure}[!t]
	\centering
	\subfloat[ Slice of a healthy sample]{\includegraphics[width=0.47\textwidth]{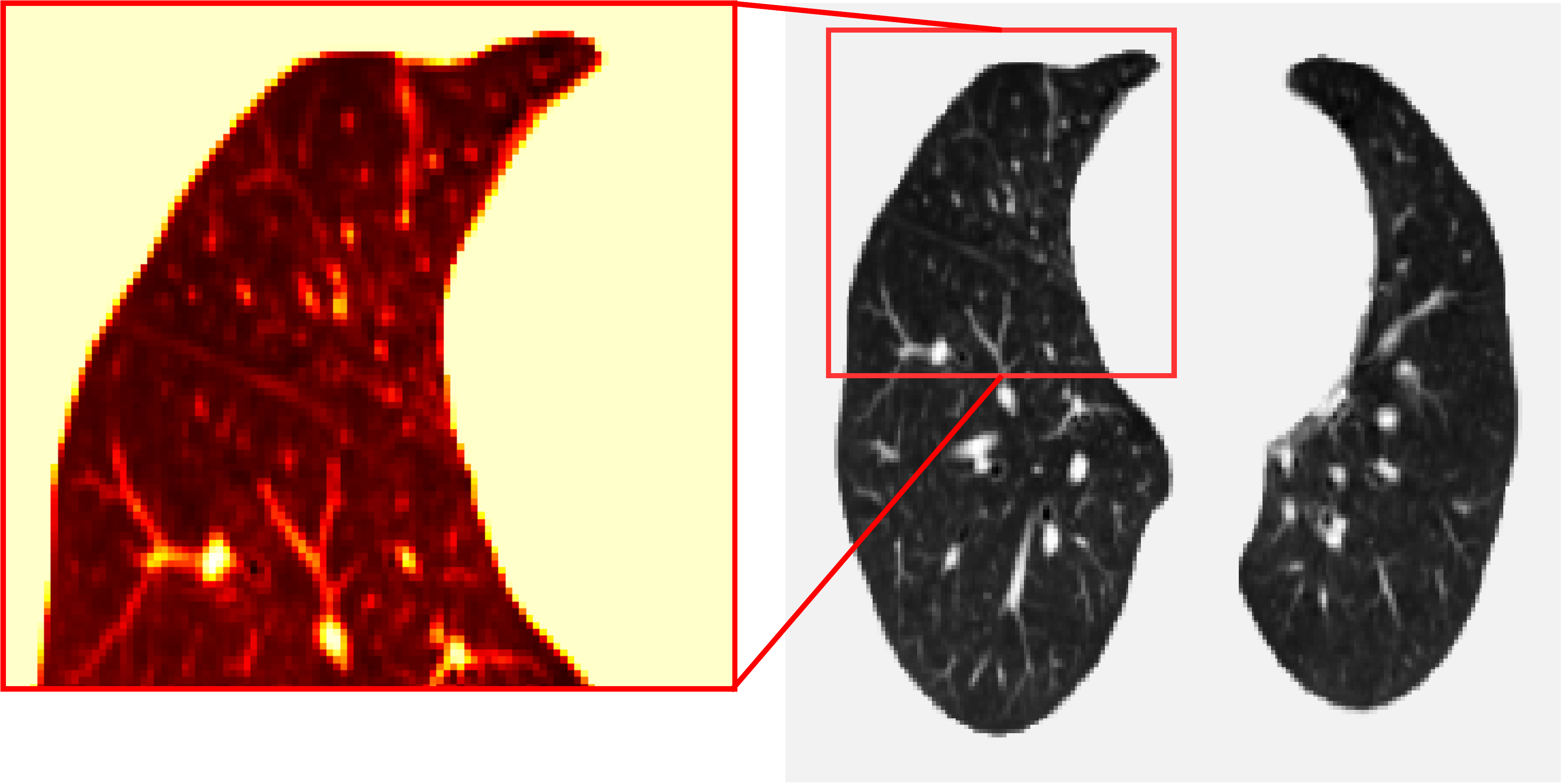}}
	\subfloat[Slice from a diseased sample]{\includegraphics[width=0.47\textwidth]{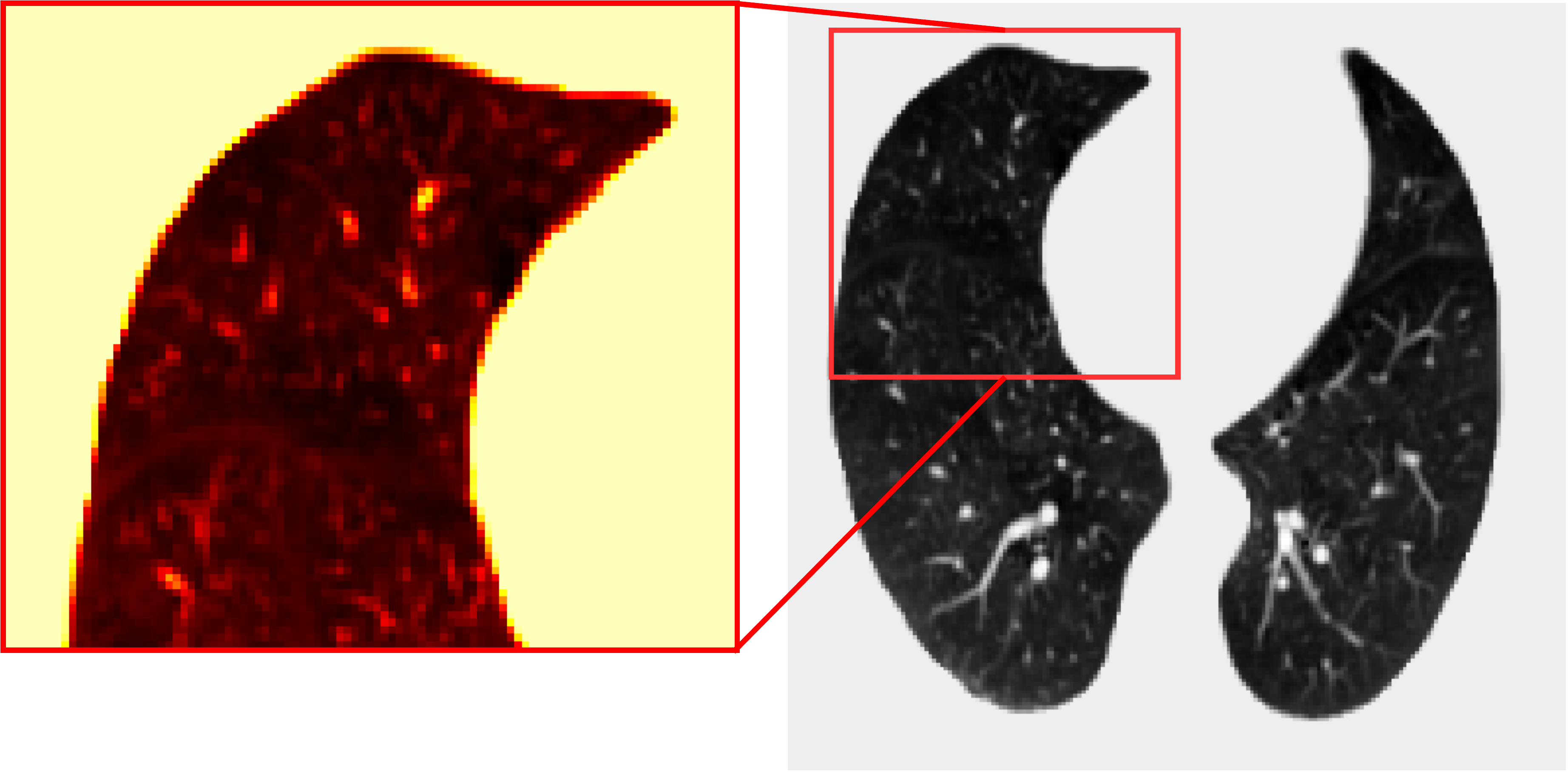}}
	\caption[SI-I samples]{The slices show more dark regions when perceived with colormap. These dark regions are an indicator of empty air space, and an abundance of these is an indicator of alveoli enlargement because of COPD.}
	\label{fig:SI_I_sampled}
\end{figure}

\textbf{Data Pre-processing.} Both data sets were available in the form of DICOM files. Binary lung masks were extracted for SI-I and SI-II using MeVisLab \cite{mevislab}. Masking and isometric re-sampling to a voxel size of 1mm\textsuperscript{3} resulted in volumes of various sizes in each dimension. Due to memory limitations, we downsampled the volumes to \(110 \times 200 \times 200 \) pixels for batch processing. The x and y planes were down-sampled using bi-linear interpolation, and the z-axis kept without changes. This is due to the fact, that COPDs could be very small and appear only on few slices, and downsampling could remove those information..  For a volume with \(N\) slices, 110 new slices were created by averaging over \(m=\frac{N}{110}\) slices in each new slice. A sample slice from SI-I data set is shown in Fig. \ref{fig:SI_I_sampled}.

\section{Experiments and Results}
We conducted several experiments for COPD and emphysema classification. COPD classification aimed to classify COPD vs. healthy, and emphysema classification aimed to distinguish emphysema from non-emphysema. Non-emphysema cases incorporated samples annotated as healthy or any other type of lung disease, such as SAD, lung cancer, or bronchitis etc.
\par
COPD classification performed using the SI-I data set. We employ undersampling to balance the data set and use 6000 samples for training, 500 samples for validation, and 300 samples for testing. A binary cross-entropy loss function is used with randomly initialized weights. The network was trained using an Adam optimizer with an initial learning rate of \(10^{-6}\), decay learning rate on plateau factor (\(\alpha\)) of \(0.9\), and a batch size of 2. It was not possible to increase the batch size because of memory limitations. The network was trained until there was no further decrease in the validation loss. The results are shown in Table \ref{table:comparison_COPD}. The table shows the accuracy achieved by our model for the task of COPD classification along with results published by Gonzalez et al \cite{Gonzalez2018}, and Hatt et al. \cite{Hatt2018}
\par
\begin{figure}[!t]
    \centering
		\begin{tikzpicture}[scale=0.7]
		\begin{axis}[%
		xtick=data,
		table/col sep=comma,
		xtick={10, 20, 30, 40, 50, 60, 70, 80, 90, 100},
		width=3in,
		grid=both,
		ylabel=Accuracy,
		xlabel=Epochs,
		ymin=0,xmin=0,
		legend pos=south west,
		legend style={nodes={scale=1.5, transform shape}}
		];
		\addplot [line width=0.5mm,red] table[x index=1,y index=6]{LilleFromScratchGraph};
		\addplot [line width=0.5mm,green] table[x index=1, y index=7]{FineTuningGraph};
		\addlegendentry{Random initialization}
		\addlegendentry{Transfer learning}
		\end{axis}
		\end{tikzpicture}
	\caption[From scratch vs transfer learning]{Validation accuracy comparison of our network between random initialization and transfer learning for emphysema vs. non-emphysema classification.}
	\label{fig:scratch_vs_fine}
\end{figure}
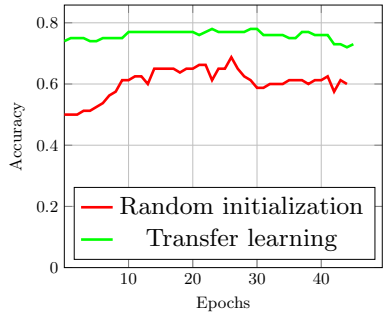

Emphysema vs. non-emphysema classification was performed using the SI-II data set. This data set was imbalanced, with 573 non-emphysema and 366 emphysema samples. The emphysema class was oversampled to balance the classes using data augmentation techniques such as flipping, rotation, and cropping. The data set was split into 80\% for training and 10\% for validation and testing each. We trained the proposed model to investigate the effect of transfer learning on the emphysema classification task. In the first experiment, we trained the model for emphysema classification using randomly initialized weights. In the second experiment, we used the pre-trained weights of our COPD classification network for weights initialization. Both networks were trained using an Adam optimizer with an initial learning rate of \(10^{-6}\), and decay learning rate on plateau factor (\(\alpha\)) of 0.9. Fig. \ref{fig:scratch_vs_fine} shows the comparison of the validation accuracy between randomly initialized weights and weight initialization using the COPD classification network using transfer learning. The results for emphysema classification are shown in Table \ref{table:comparison_emphysema}. The table shows the results for emphysema classification with and without employing transfer learning. 

\begin{table}[!t]
\caption{Results for COPD classification.}
\label{table:comparison_COPD}
\centering
\begin{tabular}{lccc}
\hline
                & \textbf{Validation}& \multicolumn{2}{c}{\textbf{Test}} \\
                & \textit{Accuracy}   & \textit{Accuracy}     & \textit{AUC}       \\
\hline
\textbf{Gonzalez et al.}\cite{Gonzalez2018}& N/A        & N/A          & 0.856     \\
\textbf{Hatt et al.}\cite{Hatt2018}     & 77.7\%       & 76.2\%         & N/A       \\
\textbf{VoxResNet 3D (ours)}   & 79.8\%       & 74.3\%         & 0.820     \\
\hline
\end{tabular}
\end{table}

\begin{table}[!t]
\caption{Results for emphysema classification.}
\label{table:comparison_emphysema}
\centering
\begin{tabular}{lccc}
\hline
& \textbf{Validation} & \multicolumn{2}{c}{\textbf{Test}} \\
& \textit{Accuracy}   & \textit{Accuracy}        \\
                                                                                                            \hline
\begin{tabular}[c]{@{}l@{}}\textbf{VoxResNet 3D} - Transfer learning\end{tabular}   & 78.3 \%       & 70.0 \%       \\
\begin{tabular}[c]{@{}l@{}}\textbf{VoxResNet 3D} - No transfer learning\end{tabular} & 68.5 \%       & 58.8 \%     \\
\hline
\end{tabular}
\end{table}

\section{Discussion}
The main focus of our work was to study the effects of using a 3D CNN for COPD classification, and effects of using transfer learning for emphysema classification. Current systems for COPD classification use CNN based on 2D kernels \cite{Hatt2018} \cite{Gonzalez2018}. While 2D CNNs have shown success in COPD classification, they intrinsically lose the 3D context \cite{7950542}. Compared to 2D, 3D kernels are able to learn discriminative features in all 3 spatial dimensions. 2D CNNs require slice-wise annotations. As diseased lung tissue is not visible in each slice of CT volume, therefore, volume-wise COPD classification is also a challenging task because in a complete CT volume annotated as COPD, there may be many slices with only healthy tissue. 
We aim to overcome these limitations by utilizing a 3D CNN for COPD classification. Our work is not directly comparable to the state-of-the-art systems because of different model parameters, training strategies, and data splits. Gonzalez et al. \cite{Gonzalez2018} extract 4 slices using anatomical landmarks to create 2D montages to train a 2D CNN. Hatt et al. \cite{Hatt2018} divides a single CT volume into 4 different 2D montages by extracting random axial slices and trains a 2D CNN. 
In comparison, we used complete CT volumes down sampled to \(110 \times 200 \times 200\) with volume-wise annotations. We achieved comparable results to \cite{Hatt2018}, \cite{Gonzalez2018} indicating that 3D CNNs show promising results with limited training data.
Table \ref{table:comparison_COPD} summarizes the results for COPD classification in comparison with Hatt et al. \cite{Hatt2018} and Gonzalez et al.\cite{Gonzalez2018}.

\par
For emphysema vs. non-emphysema classification, we compare the performance of randomly initialized weights vs. transfer learning (Fig \ref{fig:scratch_vs_fine}). As shown in Table \ref{table:comparison_emphysema}, without the use of transfer learning our network achieved 68.5\%  validation and 58.8\% test accuracy. In comparison, with transfer learning, we were able to achieve 78.3\% validation and 70.0\% test accuracy. A significant increase in performance with transfer learning indicates that features learned by neural networks for COPD classification are also effective in emphysema classification. The transfer learning step included knowledge transfer across different data sets, i.e. SI-I and SI-II, and different tasks, i.e. COPD classification and emphysema classification.

\section{Conclusion}
In this paper, we proposed a variant of 3D VoxResNet for COPD and emphysema classification. The model uses volume-wise annotations without any further feature enhancement or addition of meta-data. Our network achieved similar results for the COPD classification. For the emphysema classification, we fine-tuned the COPD classification network, which significantly increased the model performance. As a future work, validating the model on larger and balanced data sets with a thorough comparison with other methods, preferably, using k-fold cross validation is recommended. \\

\noindent\textbf{Disclaimer:} The concepts and information presented in this paper are based on research and are not commercially available.

\bibliography{main}

\begin{thebibliography}{10}

\bibitem{WHO}
World~Health Organization.
\newblock Fact sheet on chronic obstructive pulmonary disease (copd).
\newblock \url{http://www.who.int/en/news-room/fact-sheets}.

\bibitem{10.1371/journal.pmed.0030442}
C.~D. Mathers and D.~Loncar.
\newblock Projections of global mortality and burden of disease from 2002 to
  2030.
\newblock {\em PLOS Medicine}, 3(11):1--20, 11 2006.

\bibitem{doi:10.1111/j.1742-1241.2007.01447.x}
D.~Bellamy and J.~Smith.
\newblock Role of primary care in early diagnosis and effective management of
  copd.
\newblock {\em International Journal of Clinical Practice}, 61(8):1380--1389,
  2007.

\bibitem{Hatt2018}
C.~Hatt, C.~Galban, W.~Labaki, E.~Kazerooni, D.~Lynch, and M.~Han.
\newblock {Convolutional neural network based COPD and emphysema
  classifications are predictive of lung cancer diagnosis}.
\newblock In {\em Lecture Notes in Computer Science}, 2018.

\bibitem{Gonzalez2018}
G.~Gonzalez, S.~Y. Ash, G.~Vegas-S{\'{a}}nchez-Ferrero, J.~O. Onieva, F.~N.
  Rahaghi, J.~C. Ross, A.~D{\'{a}}z, R.~S.~Jos{\'{e}} Est{\'{e}}par, and G.~R.
  Washko.
\newblock {Disease staging and prognosis in smokers using deep learning in
  chest computed tomography}, 2018.

\bibitem{Ersa2015}
E.~M. {Karabulut} and T.~{Ibrikci}.
\newblock Emphysema discrimination from raw hrct images by convolutional neural
  networks.
\newblock In {\em ELECO}, pages 705--708, Nov 2015.

\bibitem{Ying2016}
J.~{Ying}, J.~{Dutta}, N.~{Guo}, L.~{Xia}, A.~{Sitek}, Q.~{Li}, and Q.~{Li}.
\newblock Gold classification of copdgene cohort based on deep learning.
\newblock In {\em ICASSP}, pages 2474--2478, 2016.

\bibitem{Chen_et_all}
H.~Chen, Q.~Dou, L.~Yu, J.~Qin, and P.~Heng.
\newblock Voxresnet: Deep voxelwise residual networks for brain segmentation
  from 3d mr images.
\newblock {\em NeuroImage}, 170:446--455, 2018.

\bibitem{mevislab}
{MeVis Medical Solutions AG}.
\newblock Mevislab.
\newblock https://www.mevislab.de/.

\bibitem{7950542}
X.~{Huang}, J.~{Shan}, and V.~{Vaidya}.
\newblock Lung nodule detection in ct using 3d convolutional neural networks.
\newblock In {\em 2017 IEEE 14th International Symposium on Biomedical Imaging
  (ISBI 2017)}, pages 379--383, April 2017.

\end{thebibliography}

\end{document}